\documentclass[aps,pre,twocolumn,floatfix,nofootinbib]
{revtex4}
\usepackage{epsfig}
\newcommand{\sgn}{\,{\rm sgn}\,}
\newcommand{\eg}{{\it e.g.\ }}
\newcommand{\ie}{{\it i.e.\ }}
\newcommand{\ignore}[1]{}

\begin{document}
\title{Topology of biological networks and
reliability of information processing}
\author{Konstantin Klemm\protect$^{\dagger\ddagger}$ and
Stefan Bornholdt\protect$^{\dagger\S}$}
\affiliation{
\protect$^\dagger$Interdisciplinary Center for Bioinformatics,
University of Leipzig, Kreuzstr.~7b, D-04103 Leipzig, Germany, 
\protect$^\ddagger$Bioinformatics Group, Dept. of Computer Science,
University of Leipzig, Kreuzstr.~7b, D-04103 Leipzig, Germany,
\protect$^\S$Institute for Theoretical Physics, 
University of Bremen, Otto-Hahn-Allee, D-28359 Bremen, Germany
}
\date{\today}
\begin{abstract}
Biological systems rely on robust internal information processing: 
Survival depends on highly reproducible dynamics of 
regulatory processes. Biological information processing elements,
however, are intrinsically noisy (genetic switches, neurons, etc.). 
Such noise poses severe stability problems to system behavior as 
it tends to desynchronize system dynamics (e.g. via fluctuating
response or transmission time of the elements). Synchronicity in
parallel information processing is not readily sustained in the absence 
of a central clock. Here we analyze the influence of topology on 
synchronicity in networks of autonomous noisy elements.  In 
numerical and analytical studies we find a clear distinction between
non-reliable  and reliable dynamical attractors, depending on the 
topology of the circuit. In the reliable cases, synchronicity is sustained, 
while in the unreliable scenario, fluctuating responses of single 
elements can gradually desynchronize the system, leading to 
non-reproducible behavior. We find that the fraction of reliable
dynamical attractors strongly correlates with the underlying circuitry.
Our model suggests that the observed motif structure of biological 
signaling networks is shaped by the biological requirement for 
reproducibility of attractors. 
\end{abstract}
\maketitle

\section{Introduction}

The processes of Life in living cells and organisms rely on highly reproducible 
information processing. It has been a long-standing question how this is 
accomplished even though it involves elements with non-reproducible, 
noisy dynamics \cite{Rao02}. In nerve cells, for instance, firing of spikes is 
not fully determined by synaptic input \cite{Allen94}. Similarly in gene regulation, 
protein concentration evolves in a relatively irreproducible manner
under given promoter levels \cite{McAdams97}. In a steady state, 
fluctuations can be dampened by the properties of the single elements as, 
\eg, autoregulation attenuates noise in genetic transcription \cite{Becskei00}. 
In a dynamical scenario with rising and falling levels of activation, intrinsic
noise of the interacting elements causes fluctuations of switching delays 
\cite{McAdams97,Beierholm01}. These cause stability problems in larger
systems of switching elements, as timing and coordination may fail. 
In order to generate reproducible dynamics, the non-deterministic responses 
of single elements have to be compensated for by a suitable circuit design. 

During recent years, biochemical interactions involved in the information 
processing within or between cells of an organism have been characterized 
systematically. For many systems of sizes up to a few hundred constituents 
a significantly large fraction of interactions have been identified such that 
the network-like structure of the system becomes visible. Such networks 
have been obtained for gene regulation \cite{Thieffry98,Costanzo01}, 
signal transduction (See for instance http://stke.sciencemag.org), 
and neuronal information processing 
\cite{White86}. Despite their different functional origins, these systems 
show strong similarity in their graph representation as nodes and interactions. 
In \cite{Milo04}, local wiring structure has been analyzed for several network 
examples in terms of abundance of connected subgraphs. 
Comparing with randomly wired graphs, real networks typically exhibit 
significant abundance of some subgraphs (called motifs), while others 
are strongly suppressed \cite{Milo02}. Networks of signal transduction and 
transcription and neural networks share similar statistics of local wiring 
patterns \cite{Milo04}. A vastly differing statistics of subgraphs is observed in 
non-signaling networks from different areas, such as the World Wide Web, 
social acquaintance webs and the graph of word-adjacency in various languages. 

In this paper, we argue that biological signaling networks (as gene networks 
or neural networks) are shaped by the common requirement of  robust signal 
processing. We study reliability of dynamics in small networks of information 
processing units with fluctuating response times. Starting with the simultaneous 
oscillation of two mutually coupled nodes we already observe that there are 
two distinct classes of dynamics. In reliable dynamics the nodes regain 
synchronization after a perturbation (retarded response) while for unreliable 
dynamics the system does not have a tendency to self-synchronize.
This distinction appears in descriptions at different levels of abstraction.
The differential equations for continuous variables (macromolecule
concentrations or firing rates, respectively) yield the same synchronization
properties as a model with discrete binary state variables. Turning to 
networks with three nodes we find that the occurrence of reliable dynamics 
is highly correlated with the underlying topology. One observes that reliable 
dynamics is more likely to appear on those triads that have been found as 
building blocks (motifs) in real biological networks. Further insight into the 
relationship between topology and reliability is gained by the analysis of 
cyclic behavior (attractors). The dynamics are reliable only if all switching 
events are connected by a single causal chain. An especially instructive 
case are isolated feedback loops. Here the fraction of initial conditions 
with reliable dynamics is obtained easily. In the following, let us start with 
the simplest example of two coupled nodes. 

\section{Feedback loops of two nodes} \label{sec:twonodes}

The distinction between reproducible and non-reproducible dynamics
becomes obvious already in the setting of two mutually coupled nodes.
Two basic scenarios are to be distinguished: (i) Nodes influence each
other with the same sign. For instance node 2 represses node 1 while
node 1 activates node 2. (ii) Both nodes have the same coupling, \ie
either both couplings are activating or both are repressing.

Starting with scenario (i), let  us describe the dynamics of the system 
by rate equations for normalized concentrations (firing rates)
$u, v \in [0,1]$ of the two nodes
\begin{eqnarray}
\label{eq:diffpmu}
\dot{u}(t) & = & \frac{\alpha_1 v^h(t-\tau_1)}{k_1^h+v^h(t-\tau_1)} -
                                                              \beta_1 u(t)\\
\label{eq:diffpmv}
\dot{v}(t) & = & \frac{\alpha_1 k_2^h}{k_2^h+u^h(t-\tau_2)} - \beta_2 v(t)~.
\end{eqnarray}
The time constants $\alpha_i$ and $\beta_i$ are the rates of
production and degradation of messenger substance, respectively.
In the neural system they represent the time constants of the varying
firing rate. The non-linear production term with Hill exponent
$h$ describes collective chemical effects or a non-linear
neural response function, respectively. Note that the rate equations
involve explicit delay times $\tau_1$ and $\tau_2$ for signal transmission. 
By randomly varying these delay variables in time, an uncertainty in 
response times of the signal processing elements is implemented.

\begin{figure}[hbt]
\centerline{\epsfig{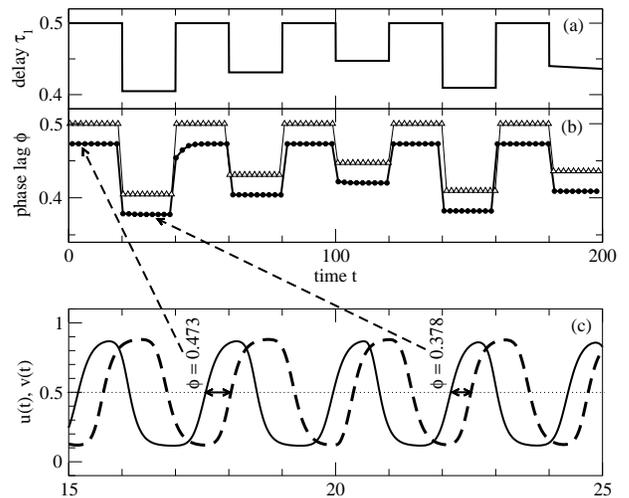}}
\caption{\label{fig:diff_2} Dynamics of the feedback loop with 
antisymmetric coupling. 
(a) Fluctuations of transmission delay $\tau_1$.
(b) Evolution of phase lag between $u(t)$ and $v(t)$ for the continuous case
(closed circles) and the discrete approximation (open triangles). Cf.\ Eqs.\
(\ref{eq:diffpmu}),(\ref{eq:diffpmv}) and
(\ref{eq:discpmu}),(\ref{eq:discpmv}), respectively.
(c) The phase lag is defined as the  time difference between subsequent
passages of variables $u$ (dashed curve)  and $v$ (solid curve) across the
value $1/2$ from below, as indicated by the black double arrows in the lower
panel. 
Parameters in Eqs.\ (\ref{eq:diffpmu}) and (\ref{eq:diffpmv}) have values 
$\tau_2=0.5$, $\alpha_1 = \alpha_2 = \beta_1 = \beta_2 = 10$,
$k_1 = k_2 = 10^{1/2}$, $h=2$. Broad variations of these parameters give
qualitatively the same results.}
\end{figure}

Figure \ref{fig:diff_2} shows time series for the above system under
varying transmission delay $\tau_1$. The variables $u(t)$ and $v(t)$ 
oscillate with a time lag that depends on the current transmission 
delay $\tau_1$. Note that (a) the dynamics remains
qualitatively the same independently of $\tau_1$, as $v(t)$ always
lags behind $u(t)$ by about $1/4$ of an oscillation period; (b) the
current phase lag does not depend on the history of the system. 
The system retains synchronization irrespective of the
changes in the transmission delay. 

\begin{figure}[hbt]
\centerline{\epsfig{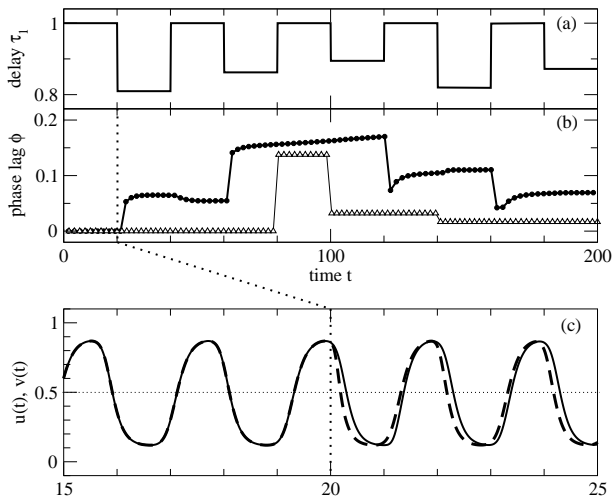}}
\caption{\label{fig:diff_3}
Dynamics of the the feedback loop with symmetric
coupling (Eqs.\ (\ref{eq:diffmmu}) and
(\ref{eq:diffmmv})) under fluctuations of the transmission delay $\tau_1$.
We keep $\tau_2=1.0$ constant, all other parameters and plotting details
are the same as in Fig.\ \protect\ref{fig:diff_2}.
}
\end{figure}

Approximating the sigmoid response curves by step functions and
considering the limit of fast production and degradation
$\alpha_i,\beta_i \rightarrow \infty$  in Eqs.\ (\ref{eq:diffpmu}) and
(\ref{eq:diffpmv}) we obtain the simplified dynamics
\begin{eqnarray}
\label{eq:discpmu}
u(t) & = & v(t-\tau_1)\\
\label{eq:discpmv}
v(t) & = & 1 - u(t-\tau_2) 
\end{eqnarray}
where now we have binary state variables $u,v \in \{0,1\}$. 
Fig. \ref{fig:diff_2} shows that this approximate description has 
qualitatively the same behavior as the original, continuous rate equations.
 Let us now turn to case (ii) above. Consider two mutually coupled
nodes with the {\em same} type of coupling in both directions. Here we take
both couplings to be inhibitory. Again we describe the system in terms of rate
equations
\begin{eqnarray}
\label{eq:diffmmu}
\dot{u}(t) & = & \frac{\alpha_1 k_1^h}{k_1^h+v^h(t-\tau_1)} - \beta_1 u(t)\\
\label{eq:diffmmv}
\dot{v}(t) & = & \frac{\alpha_1 k_2^h}{k_2^h+u^h(t-\tau_2)} - \beta_2 v(t)~.
\end{eqnarray}
The approximation by binary state variables reads
\begin{eqnarray}
\label{eq:discmmu}
u(t) & = & 1 - v(t-\tau)\\
\label{eq:discmmv}
v(t) & = & 1 - u(t-\tau)~.
\end{eqnarray}
The time series for both the continuous and the binary system are
plotted in Fig.\ \ref{fig:diff_3}. Initially, the variables $u(t)$ and
$v(t)$ are perfectly synchronized, due to identical initial values.
Changing the transmission delay $\tau_1$ causes a non-zero phase lag,
that is retained even when the transmission delay is reset to the
original value. In fact, varying $\tau_1$ inside a small interval
causes accumulation of phase lag between $u(t)$ and $v(t)$. The system
does not self-synchronize and may therefore be driven out of phase by
slight fluctuations of transmission delay.

\begin{figure}[hbt]
\centerline{\epsfig{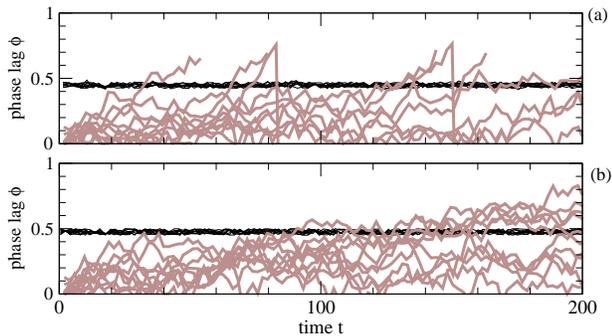}}
\caption{\label{fig:phase_0}
Systems with and without stable synchrony.
(a) Phase lag as a function of time for systems of two mutually
coupled nodes with continuous state variables under fluctuating
transmission delays. The system with antisymmetric coupling
(Eqs.\ (\ref{eq:diffpmu}) and (\ref{eq:diffpmv})) remains
synchronous with a phase lag close to $0.5$ (thin black curves).
In the system with symmetric coupling
(Eqs.\ (\ref{eq:diffmmu}) and (\ref{eq:diffmmv})) the phase lag
between the oscillating nodes is not stable against fluctuations
(thick curves).
(b) Same as in (a) for systems with discrete (Boolean)
state variables with antisymmetric coupling
(Eqs.\ (\ref{eq:discpmu}) and (\ref{eq:discpmv}))
and symmetric coupling
(Eqs.\ (\ref{eq:discmmu}) and (\ref{eq:discmmv})).}
\end{figure}

Consequently, system (i) with antisymmetric interaction and system
(ii) with symmetric interactions differ with respect to
reproducibility, see Fig.\ \ref{fig:phase_0}. With transmission delays
varying randomly with time and across several runs, system (i) shows
reproducible behavior. The phase lag practically remains constant, 
as the oscillation with a given phase lag is the stable (attractive)
mode. In contrast, the behavior of system (ii) varies across runs.
Synchronized oscillation is a marginally stable mode of this system
such that the fluctuations drive the system away from this mode.

\section{Three nodes} \label{sec:threenodes}
\begin{figure}[hbt]
\hspace*{5mm}
\epsfig{file=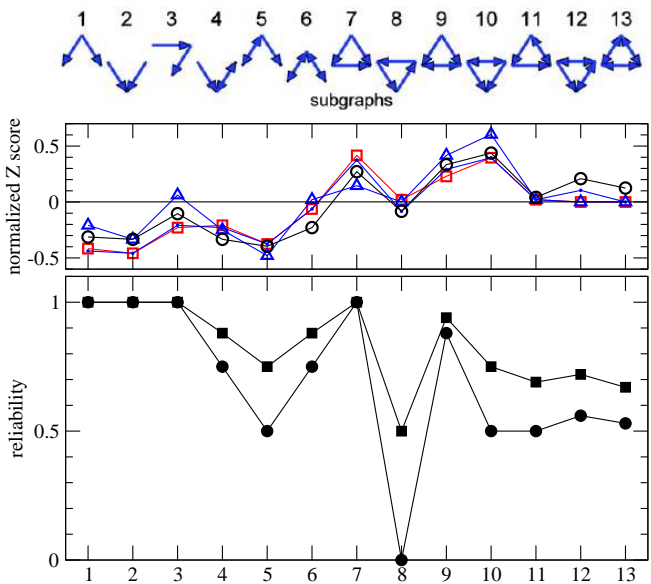,width=.44\textwidth}
\centerline{\epsfig{file=enumperfectd_1.eps,width=.48\textwidth}}
\caption{\label{fig:enum}
Abundance and dynamical reliability of subgraphs. 
Top: All directed connected graphs of three nodes.
Center: Triad significance profile \cite{Milo04} in various signaling
networks.
Bottom: Graph reliability (squares) and strict graph
reliability (circles) as defined in the main text.
Triad significance profiles in the center panel are given for the
signal-transduction in
mammalian cells (squares), genetic transcription for development in
the fruit fly (dots) and the sea-urchin (triangles), synaptic
connections in C. elegans (circles). Triad significance is the
difference between numbers of occurrence in the given network and in an
ensemble of randomly rewired surrogate networks, divided by the standard
deviation. The triad significance profile is the normalized vector of
triad significances. See reference \cite{Milo04} for details.
}
\end{figure}

In the following we shall see that the clear distinction between
reproducible dynamics with intrinsically stable synchronicity and
non-reproducible dynamics sensitive to fluctuations extends
beyond simple oscillations in systems of two nodes. Let us study
dynamics on 3-node circuits as shown in the top row of
Fig.\ \ref{fig:enum}. To this end, we first need to define
the dynamics of nodes with more than one input. Consider
a node $i$ that is directly influenced by nodes $j$ and $k$.
Restricting ourselves to the binary approximation of states from
now, node $i$ switches according to
\begin{equation} \label{eq:BoolUpdate}
x_i (t) = f_i [ x_j(t-\tau_i), x_k(t-\tau_i) ]
\end{equation}
where we define $\tau_i$ as the (time-dependent) transmission delay of
node $i$. The Boolean
function $f_i$ maps the 4 pairs of binary states $(x_j,x_k)$ to the
set $\{0,1\}$. We choose $f_i$ from the set of canalizing functions
\cite{Kauffman84}, \ie we do not use
the exclusive-or function and its negation. From the 14 canalizing
functions we further exclude those that are constant with respect to
one or both of the inputs. We are then left with $8$ Boolean functions
of two inputs (these are OR, AND and their variants generated by
negation of one or both inputs). For nodes with one input,
we allow only the non-constant Boolean functions Identity
(output $=$ input) and Negation (output $\neq$ input).

As {\em subgraph reliability} we define the probability of obtaining 
reliable dynamics when preparing a random initial condition and a 
random assignment of Boolean functions. The {\em strict subgraph 
reliability} is the probability that {\em all}
initial conditions yield reliable dynamics for a random assignment of
Boolean functions. The small system size allows us to obtain the exact
values of these quantities by full enumeration of all combinations of
initial conditions and function assignments. 
See "Methods" for details on the simulation procedure. 

Viewing the results in Fig.~\ref{fig:enum} (bottom), we note first that
feed-forward wiring (subgraphs 1, 2, 3 and 7) always gives reliable
dynamics because from any initial condition the system reaches a fixed
point after a short time. The least reproducible dynamics is obtained
for the pure feedback loop (subgraph 8) with strict reliability zero.
For an explanation see the section on ``Feedback loops of arbitrary
length''. Interestingly, the feedback loop with an additional link in
the ``opposite'' direction (subgraph 11) has a larger strict
reliability $1/2$.

Now let us compare the dynamical reliability of the subgraphs with
their abundance in biological signaling networks. Consider first
subgraphs 7--13, which contain a closed triad. Among these, subgraph 7
(the feed-forward loop) and subgraph 9 (two bidirectionally coupled
nodes receiving from a common third node) have particularly large
reliability. These subgraphs are also 
observed to be 
highly abundant in the signaling
networks. Furthermore, subgraph 12 has larger reliability than
subgraphs 11 and 13. This superiority is also reflected by abundance
measure in the networks. The large 
abundance (quantified by the Z-score measure as defined in \cite{Milo04})
of subgraph 10, however,
cannot be predicted from the dynamical reliability. For the remaining
subgraphs 1--6, wiring diagrams without closed triads, there is a
clear correlation between the empirical Z score and our measure of
dynamical reliability.

The presence or absence of a given subgraph cannot be fully explained
by the reliability measure presented here. First, the Z score tends to
increase with the number of connections in the subgraph. The bias
towards densely (but not fully) connected motifs is a consequence of
the networks' modular structure with functional clusters of nodes
\cite{Milo04b}. This property of the networks is not directly related 
to robust dynamics and therefore is not reflected in our reliability 
measure. Second, we have neglected the network
environment of the motifs, that may greatly change the dynamics. For
instance, the feed-forward subgraph 3 alone yields perfectly reliable
dynamics. The 4-node feedback loop $1 \rightarrow 2 \rightarrow 3
\rightarrow 4 \rightarrow1$, however, does not give reliable dynamics
(as shown below), even though subgraph 3 is its only motif. Third, we
have defined reliability of a subgraph by averaging over all
assignments of plausible functions. In reality, however, the nodes'
functions may be correlated with the wiring diagram. For instance,
subgraph 10 gives perfectly reliable dynamics if one assumes
antisymmetric influence (one promoter, one repressor) between the
bidirectionally linked nodes.

Despite these limitations in the present analysis, the correlation between 
abundance and reliability of subgraphs in Fig.\ \ref{fig:enum} is striking. 
In the following we gain more detailed insight into the mechanism
leading to reliability by considering attractors in feedback loops.

\section{Attractors and causality}

\begin{figure}[hbt]
\centerline{\epsfig{file=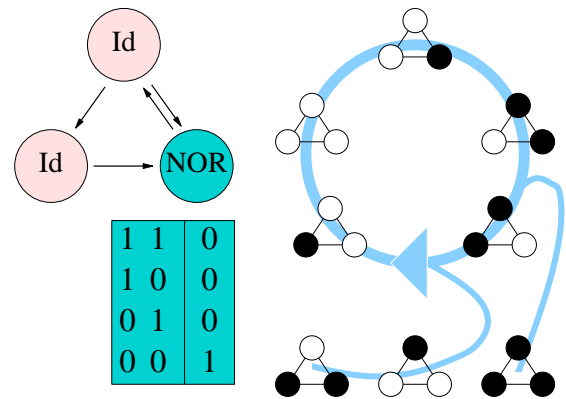,width=.41\textwidth}}
\caption{\label{fig:att13nor}
Attractor of a Boolean network. Left: A Boolean network with $N=3$
nodes with possible states $0$ (off) and
$1$ (on). Each node has a lookup table to determine its state given
the input from other nodes. Here the two light nodes simply copy their
single input, while the dark node performs the Boolean function NOR
on its inputs.
Right: In the deterministic case of constant transmission
delays, the system always reaches a periodic reproducible
sequence of states.
}
\end{figure}

Even though in biological systems transmission delays fluctuate,
it is instructive to regard constant transmission
delays as a reference case. Setting $\tau_i=1$ for all nodes
$i$ in Eq.\ (\ref{eq:BoolUpdate}), we recover the time-discrete
synchronous update mode often employed in Boolean network models.
In this idealized picture, all signal transmissions require exactly
the same time and nodes flip synchronously as if driven by a central
clock. The deterministic dynamics eventually reaches
a periodic attractor --- an indefinitely repeated sequence of
network states as illustrated in Fig.\ \ref{fig:att13nor}.

\begin{figure}[hbt]
\centerline{\epsfig{file=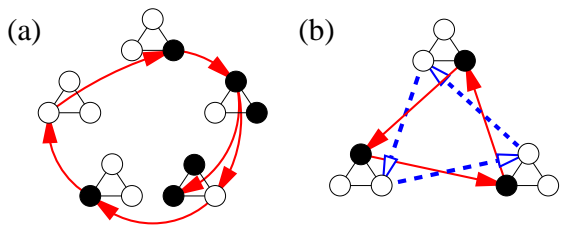,width=.41\textwidth}}
\caption{\label{fig:attcausal}
Causal structure of attractors. (a) The attractor shown in
Fig.\ \protect\ref{fig:att13nor} has one causal chain of
flipping events triggering each other. (b) A different
attractor (on the same wiring diagram) with two independent
causal chains of flipping events. The attractor in (b)
is obtained after replacing the NOR in
Fig.\ \protect\ref{fig:att13nor} with the Boolean function
that gives $1$ if and only if it receives $0$ from the top
node and $1$ from the left node.
}
\end{figure}

Reliability under fluctuating transmission delays affects the
stability of this perfectly synchronized mode. When slightly
perturbed by one retarded signal transmission, does the system
autonomously re-establish synchronicity? Or does the system
``remember'' the perturbed timing? In the latter case, just as
in Fig.\ \ref{fig:diff_3}, a series of perturbations may drive
the system away from the predictable periodic behavior.
The key to answering the question about stability is the causal
structure of the attractor. Let us call a ``flipping event''
a pair $(i,t)$, given that node $i$ changes state at time $t$.
For the attractor in Fig.\ \ref{fig:att13nor} we draw arrows
from each flipping event to all flipping events it causes to
happen in the next time step. In the resulting plot,
Fig.\ \ref{fig:attcausal}(a), there is a single closed
chain of flipping events. Retardation of one event by a time
$t_{\rm ret}$ simply retards all subsequent events by the same
amount of time. Event $(i,t)$ in the unperturbed scenario
becomes event $(i,t+t_{\rm ret})$, but the sequence of states
encountered by the system remains the same. This built-in
compensation of fluctuations renders the dynamics {\em reliable}.

An example of an {\em unreliable} dynamical attractor is shown in
Fig.\ \ref{fig:attcausal}(b). In this case there are two
separate chains of flipping events, one connecting the
on-events and the other connecting the off-events.
Retarding an event in one of the chains does not influence
the timing of the events in the other chain. By repeatedly
retarded on-events the time span a node is on is gradually reduced
and eventually reaches zero. Then the system encounters a fixed
point and does not follow the attractor any longer.
In general, the dynamics is reliable if and only if the attractor
contains exactly one causal chain.

\section{Feedback loops of arbitrary length}

\begin{figure}[hbt]
\epsfig{file=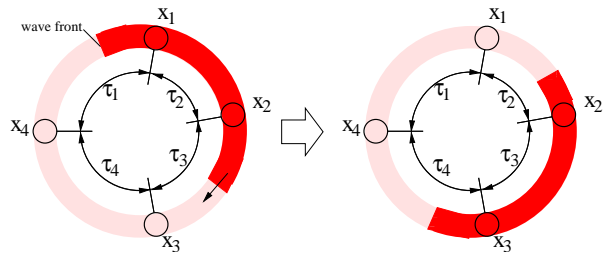,width=.44\textwidth}
\caption{\label{fig:fbloop_illu}
The ``wave front'' picture of dynamics in a feedback loop
with $N=4$ nodes all performing the Boolean function
Identity. 
}
\end{figure}

The relation between topology and reproducibility is particularly
obvious in isolated feedback loops. Consider $N\ge2$ nodes connected
in a directed cycle, \ie each node receives input only from its
(clockwise) predecessor. If node $i-1$ changes state (``flips'') at
time $t$, node $i$ will change state at time $t+\tau_i$, node $i+1$
will change state at time $t+\tau_i+\tau_{i+1}$ and so forth. The
dynamics can be interpreted as ``wave fronts'' of flipping events
traveling at constant speed on a ring. The nodes are located on this
ring at distances given by the transmission delays $\tau_i$, as
illustrated in Fig.\ \ref{fig:fbloop_illu}. For constant transmission
delays, the dynamics is periodic (with period $\tau_1+\dots+\tau_N$).
However, when a transmission delay $\tau_i$ fluctuates, consecutive
passages of wave fronts from node $i-1$ to node $i$ take different
times. Eventually one wave front may catch up with the other. Wave
fronts annihilate upon encounter: Flipping from 0 to 1 and back to 0 at
the same time results in no flipping at all. Annihilations of wave
fronts happen stochastically as they are driven by the random
fluctuations of transmission delays.

Consequently, the dynamics is reproducible only if annihilation
of wave fronts is excluded. When all nodes perform the function
Identity, as in Fig.\ \ref{fig:fbloop_illu}, the number of wave fronts is 
even. Two or more wave fronts can annihilate, eventually leading to 
irreproducible dynamics. The only reproducible dynamics in this
case is a system that stays on a fixed point, corresponding to zero  
wave fronts.

However, if one of the node performs the function Negation
then this node acts as a resting wave front, because states
on the two sides are always different. The total number of
wave fronts (including the resting one) is still even, but
now the number of {\em traveling} wave fronts is odd. Initial
conditions exist such that there is a single traveling
wave front, giving reproducible dynamics.

The two cases generalize easily. For an even number of inhibitory
couplings (\ie an even number of nodes performing Inversion) the
dynamics is reliable if and only if one of the two fixed points is
chosen as initial condition. Then the fraction of initial conditions
with reproducible dynamics is
\begin{equation}
q_{\rm even} = 2 / 2^N =2^{N-1}~.
\end{equation}
for a feedback loop of $N$ nodes.
Analogously we find
\begin{equation}
q_{\rm odd} = 2 N / 2^N = N 2^{N-1}
\end{equation}
because in a feedback loop with an odd number of Negations, there
are $2N$ initial conditions that generate a single wave front.
Note that the only feedback loop that yields reproducible dynamics
for {\em all} initial conditions has $N=2$ nodes, one performing
Inversion and the other Identity. This is the system studied as case
(i) given by Eqs.\ (\ref{eq:diffpmu}) and (\ref{eq:diffpmv}).

\section{Discussion and Conclusions} \label{sec:conclusion}

Biological information processing systems are constrained by the need for
reproducible output from circuits of intrinsically noisy elements. Compensation
for the effects of  fluctuations  on the system level  can be achieved  through
the selection of a suitable circuit design. In the present work we have
studied the influence of network topology on the reliability of information
processing by  networks of  switches with fluctuating response times.
Analytical considerations and simulation studies have shown that the dynamical
scenarios occurring in the deterministic case fall into two classes in the
presence of fluctuations. In reliable scenarios, the elements cooperatively
suppress the fluctuations and tend to synchronize their operations. Unreliable
scenarios, in contrast, do not favor synchronization and show irreproducible,
desynchronized behavior when response times fluctuate.

In general, reliable and unreliable dynamical scenarios may be distinguished by
considering the causal chains (cascades) of switching events in the network. A
dynamics that can be separated into causal chains without a common checkpoint
is unreliable. Even though they might be involving the same nodes, flipping
events on different causal chains cannot synchronize. 

The occurrence of the two dynamical classes is strongly biased by the
topology. Whether or not the system shows reliable dynamics can to a large
degree be deduced from the unlabeled wiring diagram without information on
the type of couplings and functions of switches. To capture this connection
between topology and dynamical robustness quantitatively we have defined
{\em graph reliability} as the probability to obtain reliable dynamics on
a given directed graph.

We have obtained the graph reliability for each of the 13 directed networks of
three switching elements. Comparing with empirical networks of genetic 
transcription, signal transduction, and the nervous system we find that the 
statistics of the local wiring structure  is closely related to  the reliability measure. 
Reliable triads tend to occur significantly more frequently in natural networks 
compared to the randomized versions of the networks, while unreliable
triads are typically suppressed when comparing empirical nets with their
randomized counterparts.

The fact that without knowledge of functionality  prominent features of  the
local wiring follow solely from the constraint of reliability, supports the
hypothesis that noise effects are suppressed at the system level by selection
of topology. Reversely, our approach might be used to predict function as the
most reliable dynamics given an observed wiring pattern. For instance, we may
infer that a feedback loop should contain an odd number of suppressing elements
because then the dynamics is reliable more often than in the case of even
number of suppressors.

Several extensions of this work are conceivable.  In this first dynamical study
of motifs we have neglected the network context by assuming all external input
to be constant. Further work should drop this assumption and study larger
groups of nodes containing a given subgraph to be tested.
Another interesting outlook is a  study of the dynamics on an empirical network
\cite{Kauffman03} and to compare its reliability  with rewired counterparts. We
expect that dynamical studies on biological network topologies will teach us
about origin and function of these systems,  even when still lacking full
knowledge of network properties and dynamics.

\section{Methods} \label{sec:methods}

Differential equations are integrated by first order Euler method using
a time increment $\Delta t = 10^{-5}$. For systems with Boolean variables
(Eqs.\ (\ref{eq:discpmu}), (\ref{eq:discpmv}), (\ref{eq:discmmu}),
(\ref{eq:discmmv}), and (\ref{eq:BoolUpdate}) integration is performed
with continuous time $t$ (exact up to machine precision). In the
simulations in section \ref{sec:threenodes},
transmission delays $\tau_i$ are varied as follows: Whenever node $i$
changes state, a new target delay time $\tau_i^\ast$ is drawn from the
homogeneous distribution on the interval $[0.9;1]$. Then the delay
$\tau_i$ approaches the target value $\tau_i^\ast$ according to
$\dot{\tau_i} = \Delta t \sgn(\tau_i^\ast-\tau_i)$ (if $\tau_i$ were set
to a new value abruptly, temporal order of signal transmission would not
be conserved).

Criterion for reliability: For a given dynamical scenario
(choice of Boolean functions and initial condition) 100 independent
simulation runs of duration $T=1000$ are performed and the sequence
of sustained states is recorded. A sustained state is a state
$(x_1,\dots,x_N)$ that is assumed by the system for time at least
$t_{\rm sus}=1/2$. The given dynamical scenario on the given subgraph
is called reproducible if the series of sustained states of all
100 runs are identical. Varying the number of runs and their duration
by one order of magnitude does not change the set of reliable
scenarios.

\section{Acknowledgments}
This work was supported by the Deutsche Forschungsgemeinschaft DFG.

\end{document}